\documentclass[aps,preprintnumbers,secnumarabic,amsmath,amssymb,usenatbib,nofootinbib,superscriptaddress,showkeys,showpacs,11pt]{revtex4-2}
\usepackage{graphicx}
\usepackage{dcolumn}
\usepackage{bm}
\usepackage{latexsym}
\usepackage{amsfonts}
\DeclareMathOperator{\sech}{sech}
\usepackage{amssymb}
\usepackage{amsmath}
\usepackage{subfigure}
\usepackage{mathrsfs}
\usepackage{tabu}
\usepackage[colorlinks]{hyperref}
\usepackage{orcidlink}

\begin{document}

\title{Massive Dirac particles based on gapped graphene with Rosen-Morse potential in a uniform magnetic field}

\author{A. Kalani}
\affiliation{Department of Physics, Ayatollah Amoli Branch, Islamic Azad University, Amol, Iran}

\author{Alireza Amani\orcidlink{0000-0002-1296-614X}}
\email{al.amani@iau.ac.ir}
\affiliation{Department of Physics, Ayatollah Amoli Branch, Islamic Azad University, Amol, Iran}

\author{M. A. Ramzanpour}
\affiliation{Department of Physics, Ayatollah Amoli Branch, Islamic Azad University, Amol, Iran}

\date{\today}

\begin{abstract}

We explore the gapped graphene structure in the two-dimensional plane in the presence of the Rosen–Morse potential and an external uniform magnetic field. In order to describe the corresponding structure, we consider the propagation of electrons in graphene as relativistic fermion quasi-particles, and analyze it by the wave functions of two-component spinors with pseudo-spin symmetry using the Dirac equation. Next, to solve and analyze the Dirac equation, we obtain the eigenvalues and eigenvectors using the Legendre differential equation. After that, we obtain the bounded states of energy
depending on the coefficients of Rosen–Morse and magnetic potentials in terms of quantum numbers of principal $n$ and spin–orbit $k$. Then, the values of the energy spectrum for the ground state and the first excited state are calculated, and the wave functions and the corresponding probabilities are plotted in terms of coordinates $r$. In what follows, we explore the band structure of gapped graphene by the modified dispersion relation and write it in terms of the two-dimensional wave vectors $K_x$ and $K_y$. Finally, the energy bands are plotted in terms of the wave vectors $K_x$ and $K_y$ with and without the magnetic term.

\end{abstract}



\maketitle

\section{Introduction}\label{s1}

Today, graphene is a widely used material due to its excellent electronic, optical, thermal and mechanical properties. The structure of this substance is one of the allotropes of carbon, which is formed in a two-dimensional plane of carbon in the form of a honeycomb network. Each carbon atom has three covalent bonds with three other carbon atoms and shares one free electron between all atoms. In this case, free electrons in graphene exhibit relativistic behavior and follow the Dirac equation. Therefore, for the analysis of relativistic fermions in graphene, the Dirac equation is implemented with two approaches as massless graphene and massive graphene. The mass terms in massless and massive graphenes are zero and nonzero, respectively, which means that the variation of energy has a linear and quadratic relationship in terms of momentum, respectively. In this case, the corresponding Dirac fermions have the normal and anomalous electronic properties, respectively. The noteworthy point is that in massive graphene, the Dirac fermion has a variable speed that can approach the speed of light, but this speed is reduced to one hundredth in massless graphene compared to massive state \cite{Khveshchenko-2009, Ulstrup-2014, Setare-2010}.

In order to describe the movement of graphene fermions, we need to obtain the relativistic wave equation with the Dirac equation. The solution of Dirac’s equation has certain complexities that some potentials have analytical solutions and with some other potentials the solutions are not analytical, so we can sometimes reach analytical solutions with mathematical tricks and initiatives. Many studies solved the Dirac equation analytically and quasianalytically, such as the Hulth\'{e}n potential \cite{Jian-2003, Ikhdair-2010}, the Woods–Saxon potential \cite{Guo-2005}, the Eckart potential \cite{Sari-2015}, the P\"{o}schl-Teller potential \cite{Wei-2009}, the Manning–Rosen potential \cite{Wei-2008}, the hyperbolic potential \cite{Jia-2009}, the Rosen-Morse potential \cite{Rosen-1932, Ikhdair-2013, Oyewumi-2010}, the pseudo-harmonic potential \cite{Gang-2004}, and the Morse potential \cite{Morse-1929, Zhang-2016, Berkdemir-2006, Ikhdair-2011}. These solutions were not necessarily for graphene but for a system with fermions moving under these potentials. It should be noted that the aforesaid potentials are even solved in some physical systems by the supersymmetry method, iteration method, and Nikiforov–Uvarov method, etc \cite{Berkdemir-2006, Ikhdair-2011, Bayrak-2007, Amani-2012}. If we pay attention to the root of the Dirac equation, it includes two potentials of repulsion vector $V(r)$ and attraction scalar $S(r)$, which are useful for determining spin and pseudo-spin symmetry, so that the vector potential and the scalar potential are coupled to mass and energy, respectively. These expressions exist in the Dirac Hamiltonian, which is invariant under the SU(2) algebra for the above-mentioned two symmetries. The corresponding symmetries with the nuclear shell model for nuclear physics phenomena were introduced by Refs.  \cite{Hecht-1969, Arima-1969}. This means that in spin symmetry, the difference between vector and scalar potentials is constant, while in pseudo-spin symmetry the sum of these two potentials is constant \cite{Smith-1971, Bell-1975, Ginocchio-2004}. In addition, the Dirac equation with spin symmetry and pseudo-spin symmetry has been solved with different potentials and with multiple methods \cite{Jiaa-2009, Zali-2021, Wei-2010, WeiS-2010, Qiang-2012}. 

Now, physically, we can ask how does pseudo-spin symmetry happen? To answer this question, we look at the motion of electrons inside a solid and assign them an effective mass as they move through undisturbed free space. So, we come to the conclusion that the effective mass contains a different mass,
which is called a quasi-particle. Therefore, the movement of electrons in a solid is the movement of quasi-particles with their effective mass. This issue is very applicable to the structure of graphene, which is a honeycomb network of individual carbon atoms. Therefore, electrons move in graphene as relativistic fermionic particles of carbon atoms that are located in the neighborhood of the Fermi level and at the edges of the first Brillouin zone, these points at the edges are called Dirac points. If we pay attention to the solutions of the Dirac equation without mass, we find that the conduction and valence bands will be in touch with each other at the Dirac points, or in other words it has a zero-energy gap. If we consider the corresponding graphene in a more realistic way, meaning that its quasi-particles have effective mass and are subjected to atomic potentials, we will see that an energy gap will be created between the conduction and valence bands, which is so-called the gapped graphene \cite{Pedersen-2009, Zhu-2009, Klimchitskaya-2017}.

In order to perform a complete and accurate investigation of graphene, the effects of electron–electron interaction as quasi-particle motion are very important. For this purpose, the electronic structures of gapped graphene can be investigated from the perspective of high energy physics, which is a meaningful bridge to condensed matter \cite{Novoselov-2004, Neto-2009, Nair-2008} in the sense that the quasi-particles are placed in front of a potential barrier, which prompts us to consider the corresponding system as a diatomic molecule. It should be noted that many papers have studied diatomic molecular systems in the different fields of physics using Dirac equation, Schr\"{o}dinger equation, and Klein–Gordon equation \cite{Chenaghlou-2021, Amani-2012, Onyenegecha-2021}. Therefore, with this point of view, in this paper we study the general state of the gapped graphene under a potential barrier called Rosen–Morse potential that can describe the interatomic interaction and inter-surface interaction for quasi-particles in pseudo-spin symmetry. Rosen and Morse studied the corresponding potential for the vibrations of polyatomic molecules, such as the vibration of nitrogen inside the ammonia molecule \cite{Rosen-1932}. In addition, we consider the present study in a uniform magnetic field, which causes changes in the band gap and the energy spectrum \cite{Eshghi-2016, Alimohammadian-2020, Downing-2016}. Since the Rosen–Morse potential is a potential that depends on the distance, it can be a very useful description of the intra-material interactions in the realistic graphene arrangement. Although the theoretical modeling was discussed in this research, it is expected that in a realistic graphene it will bring features regarding the interaction of electrons and network structure or between electrons and external fields, both electric and magnetic. Therefore, the potential can be designed to investigate the anharmonic nature and characteristics of these interactions. However, to make a graphene with Rosen–Morse potential, computational methods such as molecular dynamics simulation or quantum mechanical calculations are usually used to adapt the potential parameters to the specific properties and behaviors observed in graphene. Thus, it is possible to obtain the depth and width of the potential well by matching the range of the graphene network. However, theoretical models can provide a more accurate insight into the properties and behavior of a realistic graphene. Therefore, in this paper we solve analytically the corresponding Dirac equation in the presence of the
Rosen–Morse potential and the external magnetic field for a gapped graphene. In addition, the wave functions are obtained by the Legendre polynomial as a special function and then the bound energy in terms of the coefficients related to the corresponding potential and the magnetic field.

We organize the present study as follows. In Section \ref{II}, we generally present the massive Dirac equation in the presence of the scalar and vector potentials and the magnetic field. In Section \ref{III}, we will obtain the energy spectrum and wave functions by using the Legendre polynomial in an interatomic-surface interaction by the Rosen–Morse potential as pseudo-spin symmetry, as a result we calculate the energy spectrum in terms of orbit–spin quantum numbers and draw the wave functions in terms of radial coordinates. In Section \ref{IV}, we present the band structure of gapped graphene and draw the energy bands in terms of wave vectors. Finally, we provide a summary of the present work in Section \ref{V}.


\section{Massive Dirac equation with an external magnetic field}\label{II}

In this section, we intend to consider the motion of an electron with mass $M$ with the Fermi velocity, $v_F\simeq 10^6 \,m/s$, under a scalar potential, $S(r)$, and a vector potential, $V(r)$, in an external uniform magnetic field, $\vec{B} = B_0 \hat{z}$, with magnetic vector potential, $A$, in graphene. In that case, we can write the Dirac equation with the linear momentum operator, $\vec{p}$, in the following Hamiltonian form:
\begin{equation}\label{diraceq1}
H= v_F \,\vec{\alpha} . (\vec{p} - \frac{e}{c} \vec{A}) + \beta \left(M c^2 + S(r)\right) + V(r),
\end{equation}
where
\begin{equation}\label{alphabeta1}
\alpha= \left( \begin{array}{cc}
0 & \sigma \\
\sigma & 0 \end{array} \right), ~~~~~ \beta= \left( \begin{array}{cc}
I & 0 \\
0 & -I \end{array} \right),
\end{equation}
are $4 \times 4$ matrices in which $\sigma$ and $I$ are respectively Pauli matrices and unitary matrix as below:
\begin{equation}\label{pauli1}
\sigma_ x= \left( \begin{array}{cc}
0 & 1 \\
1 & 0 \end{array} \right), ~~~ \sigma_y= \left( \begin{array}{cc}
0 & -i \\
i & 0 \end{array} \right), ~~~ \sigma_z= \left( \begin{array}{cc}
1 & 0 \\
0 & -1 \end{array} \right), ~~~ I = \left( \begin{array}{cc}
1 & 0 \\
0 & 1 \end{array} \right).
\end{equation}

Notice that in the above Dirac equation, the radial coordinate $r$ is on the two-dimensional plane of $x-y$. Now the Dirac Hamiltonian \eqref{diraceq1} is written as the below equation
\begin{equation}\label{HpsiI}
H \Psi(r, \phi) = E \Psi(r, \phi),
\end{equation}
where $E$ is the eigenvalue, and $\Psi(r, \phi) = {\Psi_I(r, \phi) \choose \Psi_{II}(r, \phi)}$ is the wave function which one split to two-component spinors as $\Psi_I$ and $\Psi_{II}$. To insert Eqs. \eqref{diraceq1} and \eqref{alphabeta1} into Eq. \eqref{HpsiI}, we have
\begin{subequations}\label{psiIII}
\begin{eqnarray}
\Psi_{I} = \frac{v_F \, \sigma . \left(p - \frac{e}{c} A\right)}{E - M c^2 - S(r) - V(r)}\Psi_{II},\label{psiIII-1}\\
\Psi_{II} =  \frac{v_F \, \sigma . \left(p - \frac{e}{c} A\right)}{E + M c^2 + S(r) - V(r)}\Psi_{I},\label{psiIII-2}
\end{eqnarray}
\end{subequations}
where the solution of the pure Dirac equation is obtained by omitting the scalar potential, the vector potential, and the magnetic vector potential in the form
\begin{equation}\label{diraceq2}
 E^\pm = \pm \sqrt{M^2 c^4 + v_F^2 p^2},
\end{equation}
but the solution of the above Dirac equation is laborious in the existence of the corresponding potentials, however, we obtain dispersion relation by using $\sigma . a ~\sigma . b = a.b+i \sigma [a\times b]$  as follows
\begin{equation}\label{diraceq2}
 \left(E-M c^2 -S-V\right)\left(E+M c^2 +S-V\right) = v_F^2 \left(\vec{p} - \frac{e}{c} \vec{A}\right)^2.
\end{equation}

Now, in order to calculate wave function and eigenvalues of the corresponding system, we separate two-component spinor $\Psi_I$ and $\Psi_{II}$ in terms of coordinates $r$ and $\phi$ in the following form \textcolor{red}{\cite{Gupta-2008}}
\begin{equation}\label{psiIV}
{\Psi_I(r, \phi) \choose \Psi_{II}(r, \phi)} = {\psi_1(r) \, e^{i k \phi} \choose i \, \psi_2(r) \, e^{i (k + 1) \phi}},
\end{equation}
where $\psi_1$ and $\psi_2$ are the radial part of the wave functions, and $k$ is introduced spin-orbit quantum number which is a constant. Since we will consider the motion of a particle in graphene to be two-dimensional, so we convert the polar coordinate $r-\phi$ as the Cartesian coordinates $x-y$ as,
\begin{subequations}\label{rphi2xy1}
\begin{eqnarray}
& \frac{\partial}{\partial x}=\cos\phi \frac{\partial}{\partial r}-\frac{sin \phi}{r}\frac{\partial}{\partial \phi},\label{rphi2xy1-1}\\
& \frac{\partial}{\partial y}=sin \phi \frac{\partial}{\partial r}+\frac{cos \phi}{r}\frac{\partial}{\partial \phi},\label{rphi2xy1-2}
\end{eqnarray}
\end{subequations}
and on the other, we have according to coordinates $x$ and $y$ as
\begin{equation}\label{sigmap1}
\vec{\sigma}\cdot\vec{p}=\sigma_{x}p_{x}+\sigma_{y}p_{y},
\end{equation}
where $p_{x} = -i \hbar \frac{\partial}{\partial x}$ and $p_{y} = -i\hbar\frac{\partial}{\partial y}$. The magnetic vector potential, $A$, for a charged particle moving in a constant magnetic field, $\vec{\nabla} \times \vec{A} = \vec{B} = B_0 \hat{z}$, to be
\begin{equation}\label{magvecpot1}
\vec{A} = \frac{1}{2} \left(\vec{B} \times \vec{r}\right).
\end{equation}

By substituting Eqs. \eqref{psiIV}-\eqref{magvecpot1} into \eqref{HpsiI}, we obtain two coupled first-order differential equations for $\psi_1$ and $\psi_2$ in the following form
\begin{subequations}\label{psi12I}
\begin{eqnarray}
& \left(\widetilde{m} - \widetilde{E}+ \widetilde{S} +\widetilde{V}\right) \psi_1 + \left(\frac{d}{dr} +\frac{k+1}{r} - \eta r\right) \psi_2 = 0,\label{psi12I-1}\\
& \left(\widetilde{m} + \widetilde{E} + \widetilde{S} - \widetilde{V}\right) \psi_2 + \left(\frac{d}{dr} - \frac{k}{r} + \eta r\right) \psi_1 = 0,\label{psi12I-2}
\end{eqnarray}
\end{subequations}
where $\widetilde{E} = \frac{E}{v_F \hbar}$, $\widetilde{m} = \frac{M c^2}{v_F \hbar}$, $\widetilde{V} = \frac{V}{v_F \hbar}$, $\widetilde{S} = \frac{S}{v_F \hbar}$, and $\eta = \frac{e B_0}{2 c \hbar}$. The wave functions $\psi_1$ and $\psi_2$ are written together as
\begin{subequations}\label{psi12II}
\begin{eqnarray}
\psi_1 = \frac{1}{\widetilde{E} - \widetilde{m} - U(r)} \left(\frac{d}{dr} + \frac{k+1}{r} - \eta r\right) \psi_2,\label{psi12II-1}\\
\psi_2 = \frac{1}{-\widetilde{E} - \widetilde{m} + W(r)} \left(\frac{d}{dr} - \frac{k}{r} + \eta r\right) \psi_1,\label{psi12II-2}
\end{eqnarray}
\end{subequations}
where $U(r) = \widetilde{V}(r) + \widetilde{S}(r)$ and $W(r) = \widetilde{V}(r) - \widetilde{S}(r)$. We write two second-order differential equations from Eqs. \eqref{psi12II} in the following form
\begin{subequations}\label{psi12III}
\begin{eqnarray}
&\frac{d^2 \psi_1}{d r^2} + \left(\frac{1}{\widetilde{E}+\widetilde{m}-W}\frac{d W}{d r} + \frac{1}{r}\right) \frac{d \psi_1}{d r} + \bigg(\eta+ \frac{k}{r^2} \notag \\
&+(\eta r -\frac{k}{r}) \left(\frac{1}{\widetilde{E}+\widetilde{m}-W}\frac{d W}{d r}+\frac{k+1}{r}-\eta r\right)+ \big(\widetilde{E}+\widetilde{m}-W\big) \big(\widetilde{E}-\widetilde{m}-U\big) \bigg) \psi_1 = 0,\label{psi12III-1}\\
&\frac{d^2 \psi_2}{d r^2} + \left(\frac{1}{\widetilde{E}-\widetilde{m}-U}\frac{d U}{d r} + \frac{1}{r}\right) \frac{d \psi_2}{d r} + \bigg(-2 \eta \notag \\
&+(-\eta r +\frac{k+1}{r}) \left(\frac{1}{\widetilde{E}-\widetilde{m}-U}\frac{d U}{d r}-\frac{k+1}{r}+\eta r\right)+ \big(\widetilde{E}+\widetilde{m}-W\big) \big(\widetilde{E}-\widetilde{m}-U\big) \bigg) \psi_2 = 0,\label{psi12III-2}
\end{eqnarray}
\end{subequations}

Since we intend to describe gapped graphene in the presence of external electric and magnetic fields by the Dirac equation, then we have to consider the corresponding solution as the pseudo-spin symmetry. It should be noted that the Dirac equation has solutions of spin symmetry and pseudo-spin symmetry, in which $W(r) = C_s = constant$ is called spin symmetry and $U(r) = C_{ps} = constant$ is called pseudo-spin symmetry \cite{Gupta-2008, Hassanabadi-2012, Arda-2015, Min-2008, Jose-2009, Tuan-2014}. Therefore, we have $k (k+1) = l (l+1)$ and $k (k-1) = \widetilde{l} (\widetilde{l}+1)$ in which $l$ and $\widetilde{l}$ are orbital angular momentum for spin symmetry and pseudo-spin symmetry, respectively. Herein, the total angular momentum is introduced as $j = l + s$ and $\widetilde{j} = \widetilde{l} + \widetilde{s}$ for spin symmetry and pseudo-spin symmetry, respectively, in which $s = \widetilde{s} = \pm \frac{1}{2}$. We will have the corresponding relationships for spin symmetry
\begin{subequations}\label{spinsym}
\begin{eqnarray}
\textrm{aligned spin}: k = -(l+1),\,\, j = l + \frac{1}{2},\,\, k < 0,\\
\textrm{unaligned spin}: k = +l,\,\, j = l - \frac{1}{2},\,\, k > 0,
\end{eqnarray}
\end{subequations}
where $k = +1, \pm 2, \pm 3, \cdot \cdot \cdot$. But for pseudo-spin symmetry yields
\begin{subequations}\label{psespinsym}
\begin{eqnarray}
\textrm{aligned spin}: k = -\widetilde{l},\,\, j = \widetilde{l} - \frac{1}{2},\,\, k < 0,\\
\textrm{unaligned spin}: k = \widetilde{l} + 1,\,\, j = \widetilde{l} + \frac{1}{2},\,\, k > 0,
\end{eqnarray}
\end{subequations}
where $k = -1, \pm 2, \pm 3, \cdot \cdot \cdot$.

Therefore, since the spin of electrons in graphene play the role of pseudo-spin symmetry, so in this work we consider Dirac equation for pseudo-spin symmetry. In this case, Eq. \eqref{psi12III-2} becomes:
\begin{eqnarray}\label{psi12IV}
\frac{d^2 \psi_2}{d r^2} + \frac{1}{r} \frac{d \psi_2}{d r} + \left(-\eta^2 r^2 + 2 \eta k + \left(\widetilde{E}+\widetilde{m}-W\right) \left(\widetilde{E}-\widetilde{m}-C_{ps}\right) - \frac{(k+1)^2}{r^2} \right) \psi_2 = 0,
\end{eqnarray}

In the next section, we obtain the eigenvalues and the eigenvectors for the present system with a known potential instead of $W(r)$.
\section{Energy spectrum for a Rosen-Morse potential}\label{III}

In this section, we intend to study the effects of external electric and magnetic fields in gapped graphene by Rosen–Morse potential. For this purpose, we have to examine the effective mass of charge carriers (the propagated electrons) as relativistic fermionic quasi-particles through the graphene hexagonal lattice. Now, in order to solve the corresponding system, we must use the Dirac equation instead of Schr\"{o}dinger equation from the point of view of  symmetry pseudo-spin symmetry. In that case, instead of the summation of the scalar and vector potentials, we consider the Rosen–Morse potential in the following form \cite{Rosen-1932}
\begin{equation}\label{RS1}
W(r) = -V_1 \sech^2(\gamma r) + V_2 \tanh(\gamma r),
\end{equation}
where $V_1$ and $V_2$ are the depth of the potential, and $\gamma$ is the range of the potential. The Rosen-Morse potential has a minimum in equilibrium distance (bond length) $r_e = -\frac{1}{\gamma} \mathrm{arctanh}\left(\frac{V_2}{2 V_1}\right)$. Substituting Eq. \eqref{RS1} into Eq. \eqref{psi12IV} we have
\begin{eqnarray}\label{psi12V}
&\frac{d^2 \psi_2(r)}{d r^2} + \frac{1}{r} \frac{d \psi_2(r)}{d r} + \Big(-\eta^2 r^2 + 2 \eta k - \frac{(k+1)^2}{r^2}\notag \\
&+ \left(\widetilde{E}+\widetilde{m}+V_1 \sech^2(\gamma r) - V_2 \tanh(\gamma r)\right) \left(\widetilde{E}-\widetilde{m}-C_{ps}\right) \Big) \psi_2(r) = 0,
\end{eqnarray}
where by changing variable $z = \tanh(\gamma r)$ and applying it to the above equation, we will have:
\begin{eqnarray}\label{psi12VI}
& (1-z^2) \frac{d^2 \psi_2(z)}{d z^2}-2 z \frac{d \psi_2(z)}{d z} + \frac{1}{\mathrm{arctanh}(z)}\frac{d \psi_2(z)}{d z}
+\Big(-\frac{\eta^2}{\gamma^4} \frac{\mathrm{arctanh}^2(z)}{1-z^2} + \frac{2 \eta k}{\gamma^2} \frac{1}{1-z^2} \notag\\
&+ \left(\frac{(\widetilde{E}+\widetilde{m})}{\gamma^2}  \frac{1}{1-z^2} +\frac{V_1}{\gamma^2} - \frac{V_2}{\gamma^2} \frac{z}{1-z^2}\right) \left(\widetilde{E}-\widetilde{m}-C_{ps}\right) - \frac{ (k+1)^2}{(1-z^2)\, \mathrm{arctanh}^2(z)} \Big) \psi_2(z) = 0,
\end{eqnarray}

The analytical solution of the above equation is very complicated, so we can solve it by using the approximation method. Commonly in physics problems, Taylor series is used as a powerful mathematical technique to simplify models when they are impossible to achieve. This series allows us to approximate the function near that point using a polynomial. For this purpose, we approximate trigonometric terms around $z_e = \tanh(\gamma r_e) = -\frac{V_2}{2 V_1}$ in which $r_e$ is the bond length. In that case, we have
\begin{subequations}\label{appr1}
\begin{eqnarray}
&\frac{1}{\mathrm{arctanh}(z)} \simeq \frac{z_e^2 + z_e (1-2 z_e^2)\, \mathrm{arctanh}(z_e) + (1-z_e^2)^2\, \mathrm{arctanh}^2(z_e)}{(1-z_e^2)^2\, \mathrm{arctanh}^3(z_e)} - \frac{2 z_e + (1-3 z_e^2) \mathrm{arctanh}(z_e)}{(1-z_e^2)^2\, \mathrm{arctanh}^3(z_e)} z \notag\\
&+ \frac{1 - z_e \, \mathrm{arctanh}(z_e)}{(1-z_e^2)^2\, \mathrm{arctanh}^3(z_e)} z^2 + O(z)^3,\\\label{appr1-1}
&\frac{\mathrm{arctanh}^2(z)}{1-z^2} \simeq \frac{z_e^2 - 2 z_e (1-4 z_e^2) \mathrm{arctanh}(z_e) + (1-3 z_e^2+6 z_e^4)\, \mathrm{arctanh}^2(z_e)}{(1-z_e^2)^3} \notag\\
&+ \frac{-2 z_e + 2 (1-7 z_e^2)\, \mathrm{arctanh}(z_e)-8 z_e^3\, \mathrm{arctanh}^2(z_e)}{(1-z_e^2)^3} z + \frac{1 + 6 z_e\, \mathrm{arctanh}(z_e)+(1+3 z_e^2)\, \mathrm{arctanh}^2(z)}{(1-z_e^2)^3} z^2 + O(z)^3,\\\label{appr1-1}
&\frac{1}{(1-z^2)\, \mathrm{arctanh}^2(z)} \simeq \frac{3 z_e^2 + 2 z_e (1-4 z_e^2)\, \mathrm{arctanh}(z_e) + (1-3 z_e^2+6 z_e^4)\, \mathrm{arctanh}^2(z_e)}{(1-z_e^2)^3\, \mathrm{arctanh}^4(z_e)} \notag\\
&+ \frac{6 z_e + 2 (1-7 z_e^2)\, \mathrm{arctanh}(z_e) + 8 z_e^3 \, \mathrm{arctanh}^2(z_e)}{(1-z_e^2)^3\, \mathrm{arctanh}^4(z_e)} z + \frac{3 - 6 z_e \, \mathrm{arctanh}(z_e)+(1+3 z_e^2)\, \mathrm{arctanh}^2(z_e)}{(1-z_e^2)^3\, \mathrm{arctanh}^4(z_e)} z^2 + O(z)^3.\label{appr1-1}
\end{eqnarray}
\end{subequations}

On the other hand, we can respectively write the above approximation terms as
\begin{subequations}\label{appr2}
\begin{eqnarray}
&\frac{1}{\mathrm{arctanh}(z)}={a_0+a_1 z+a_2 z^2}\\\label{appr2-1}
&\frac{\mathrm{arctanh}^2(z)}{1-z^2}=b_0+b_1 z+b_2 z^2\\\label{appr2-2}
&\frac{1}{(1-z^2)\, \mathrm{arctanh}^2(z)}=c_0+c_1 z+c_2 z^2.\label{appr2-3}
\end{eqnarray}
\end{subequations}

Now we can write down coefficients $a_0$ to $c_2$ by using Eqs. \eqref{appr1} and \eqref{appr2} in the following form
\begin{subequations}\label{appr3}
\begin{eqnarray}
&a_0 = \frac{z_e^2 + z_e (1-2 z_e^2)\, \mathrm{arctanh}(z_e) + (1-z_e^2)^2\, \mathrm{arctanh}^2(z_e)}{(1-z_e^2)^2\, \mathrm{arctanh}^3(z_e)}, \\\label{appr3-1}
&a_1 = - \frac{2 z_e + (1-3 z_e^2) \mathrm{arctanh}(z_e)}{(1-z_e^2)^2\, \mathrm{arctanh}^3(z_e)},\\\label{appr3-2}
&a_2 = \frac{1 - z_e \, \mathrm{arctanh}(z_e)}{(1-z_e^2)^2\, \mathrm{arctanh}^3(z_e)},\\\label{appr3-3}
&b_0 = \frac{z_e^2 - 2 z_e (1-4 z_e^2) \mathrm{arctanh}(z_e) + (1-3 z_e^2+6 z_e^4)\, \mathrm{arctanh}^2(z_e)}{(1-z_e^2)^3}, \\\label{appr3-4}
&b_1 = \frac{-2 z_e + 2 (1-7 z_e^2)\, \mathrm{arctanh}(z_e)-8 z_e^3\, \mathrm{arctanh}^2(z_e)}{(1-z_e^2)^3},\\\label{appr3-5}
&b_2 = \frac{1 + 6 z_e\, \mathrm{arctanh}(z_e)+(1+3 z_e^2)\, \mathrm{arctanh}^2(z_e)}{(1-z_e^2)^3},\\\label{appr3-6}
&c_0 = \frac{3 z_e^2 + 2 z_e (1-4 z_e^2)\, \mathrm{arctanh}(z_e) + (1-3 z_e^2+6 z_e^4)\, \mathrm{arctanh}^2(z_e)}{(1-z_e^2)^3\, \mathrm{arctanh}^4(z_e)},\\\label{appr3-7}
&c_1 = \frac{6 z_e + 2 (1-7 z_e^2)\, \mathrm{arctanh}(z_e) + 8 z_e^3 \, \mathrm{arctanh}^2(z_e)}{(1-z_e^2)^3\, \mathrm{arctanh}^4(z_e)},\\\label{appr3-8}
&c_2 = \frac{3 - 6 z_e \, \mathrm{arctanh}(z_e)+(1+3 z_e^2)\, \mathrm{arctanh}^2(z_e)}{(1-z_e^2)^3\, \mathrm{arctanh}^4(z_e)},\label{appr3-9}
\end{eqnarray}
\end{subequations}
where by considering the value of equilibrium distance (bond length), $r_e = 2.197224577 \,fm$, we can obtain coefficients $a_0$ to $c_2$ as
\begin{subequations}\label{appr4}
\begin{eqnarray}
a_0 = 5.974907579, ~~~ a_1 = -12.19886077, ~~~ a_2 = 7.780005050,\\
b_0 = 0.2579095562, ~~~ b_1 = -1.289505649, ~~~ b_2 = 2.496412480,\\
c_0 = 23.32181071, ~~~ c_1 = -63.94181239 ~~~ c_2 = 24.53113834.
\end{eqnarray}
\end{subequations}

In what follows, we will solve the corresponding system by using the separation method. For this purpose, we can write down the wave function in terms of multiplied between an arbitrary function, $R(z)$, and the Legendre polynomial, $P_n(z)$, in the form
\begin{equation}\label{psi2}
\psi_2(z) = R(z) P_n(z),
\end{equation}
where the Legendre polynomial is an orthogonal polynomial of degrees $n$, and the its differential form is as
\begin{eqnarray}\label{Jacob1}
& (1-z^2){P_n}''(z)-2 z{P_n}'(z) + n(n+1) P_n(z)=0,
\end{eqnarray}
where one is defined over the interval [-1, 1]. Also, we can write down its Rodrigues Formula as below
\begin{eqnarray}\label{Jacob2}
&P_n(z) = \frac{1}{2^n n!} \frac{d^{n}}{d z^{n}}\left((z^2-1)^{n}\right).
\end{eqnarray}

To insert Eqs. \eqref{appr2} and \eqref{psi2} into Eq. \eqref{psi12VI}, one yields
\begin{eqnarray}\label{psi3}
&(1-z^2) {P_{n}}''(z) +\left[2 (1-z^2) \frac{R'}{R}-2 z+ {a_0+a_1 z+a_2 z^2}\right] {P_{n}}'(z) \notag \\
&\Big[(1-z^2) \frac{R''}{R}-2z \frac{R'}{R}+(a_0+a_1 z+a_2 z^2) \frac{R'}{R}-\frac{\eta^2}{\gamma^4}(b_0+b_1 z+b_2 z^2)+\frac{2 \eta k}{\gamma^2} \frac{1}{1-z^2}\notag \\
&+\left(\frac{(\widetilde{E}+\widetilde{m})}{\gamma^2}  \frac{1}{1-z^2} +\frac{V_1}{\gamma^2} - \frac{V_2}{\gamma^2} \frac{z}{1-z^2}\right) \left(\widetilde{E}-\widetilde{m}-C_{ps}\right) - (k+1)^2 (c_0+c_1 z+c_2 z^2)\Big] {P_{n}}(z)=0,
\end{eqnarray}
where index $'$ is the derivative with respect to $z$.

In order to obtain the form of function $R(z)$, we consider the second term of Eq. \eqref{psi3} to be equivalent to the second term of Eq. \eqref{Jacob1}. In that case, we have
\begin{eqnarray}\label{R1}
R(z) = R_0 \left(z -1\right)^{\frac{a_0+a_1+a_2}{4}} \left(z +1\right)^{\frac{-a_0+a_1-a_2}{4}} {e}^{\frac{a_2}{2}z},
\end{eqnarray}
where $R_0$ is introduced as an integral constant. The important feature of this solution is that the boundary condition for wave function is as $\psi(r \rightarrow 0) \neq 0$ and $\psi(r \rightarrow \infty) = 0$, i.e., the particle is present in $r \rightarrow 0$ but not in $r \rightarrow \infty$. By inserting Eq. \eqref{R1} into the third term of Eq. \eqref{psi3}, and then by equating the third term of Eq. \eqref{psi3} with the third term of Eq. \eqref{Jacob1}, the following results are obtained
\begin{subequations}\label{coef1}
\begin{eqnarray}
&\eta^2 = \frac{\gamma^4}{4} \left(\frac{a_2^2}{2 b_1}+ \frac{a_2(a_1-2)}{b_1}\right) -  \frac{\gamma^4}{2} (k+1)^2 \left(\frac{c_2}{b_2} + \frac{c_1}{b_1} \right),\label{coef1-1}\\
&4 n (n+1)+4 c_0 \,(k+1)^{2}-a_1^{2}-a_2^{2}+2 (a_1-a_0 a_2)-\frac{4 V_1}{\gamma^{2}} \left(\widetilde{E}-\widetilde{m}-C_{ps}\right)+\frac{4 b_0}{\gamma^{4}} \eta^{2} = 0,\label{coef1-2}\\
&\frac{4}{\gamma^{2}} \left(\widetilde{E} + \widetilde{m} + V_2 \right) \left(\widetilde{E}-\widetilde{m}-C_{ps}\right) + \frac{8 k}{\gamma^{2}}\eta- \left(a_0 -a_1+ a_2\right)^2 = 0,\label{coef1-3}\\
&\frac{4}{\gamma^{2}} \left(\widetilde{E} + \widetilde{m} - V_2 \right) \left(\widetilde{E}-\widetilde{m}-C_{ps}\right) +\frac{8 k}{\gamma^{2}}\eta-\left(a_0 +a_1+ a_2 \right)^2 = 0,\label{coef1-4}
\end{eqnarray}
\end{subequations}
where first two relations show the quantization of the magnetic field, i.e., $\eta$ is written in terms of $k$. Also, we obtain the energy spectra, $\widetilde{E} \equiv \widetilde{E}_{n k}$, from the combination of the last three relations in the following form
\begin{eqnarray}\label{bs1}
&\widetilde{E}_{n k} = \frac{\gamma^2}{8 V_1} \Big(4 n(n+1) + 4 c_0 (k+1)^2 + \frac{4 b_0}{\gamma^4}\eta^2- (2 a_0 a_2+a_1^2+a_2^2-2 a_1)\Big)\\ \notag
& -  \frac{\gamma^2 a_1}{4 V_2} \left(a_0+a_2\right) + C_{ps} + \widetilde{m},
\end{eqnarray}
where in order to eliminate $\eta$ by inserting Eq. \eqref{coef1-1} into Eq. \eqref{bs1} we have
\begin{eqnarray}\label{bs2}
&\widetilde{E}_{n k} = \frac{\gamma^2}{8 V_1} \Big(4 n(n+1) + 4 (k+1)^2 \big(2 c_0-\frac{b_0 c_1}{b_1}-\frac{b_0 c_2}{b_2}\big) + b_0 \big(\frac{a_2^2}{2 b_1}+ \frac{a_2(a_1-2)}{b_1}\big) \\ \notag
&- (2 a_0 a_2+a_1^2+a_2^2-2 a_1)\Big) -  \frac{\gamma^2 a_1}{4 V_2} \left(a_0+a_2\right) + C_{ps} + \widetilde{m},
\end{eqnarray}
where $\widetilde{E}_{n k}$ is dependent on the Rosen-Morse and external magnetic potentials, and is written in terms of the quantum numbers and other coefficients.
\begin{table}[h]
\caption{The energy spectrum $\widetilde{E}_{n k}$ by the values of $r_e = 2.197224577 \,fm$, $\widetilde{m} = 1 \,fm^{-1}$, $\gamma = 0.25 \,fm^{-1}$, $V_1 = -1 \,fm^{-1}$, $V_2 = 1 \,fm^{-1}$, and $C_{PS} = -6 \, fm^{-1}$ \cite{Ikhdair-2013}.} 
\centering 
\begin{tabular}{|| c | c | c | c | c | c | c  | | c | c | c | c | c | c | c ||} 
\hline\hline 
\,$n$\, & \,\,\,$k$ & $\,\widetilde{l}\,$ & \,\,$\widetilde{j}$\,\, & \,$l$\, & $n L_{\widetilde{j}}$~;~$(n-1)L_{\widetilde{j}}$ & $\widetilde{E}_{n k}\,(fm^{-1})$ & \,$n$\, & \,\,$k$ & $\,\widetilde{l}\,$ & \,\,$\widetilde{j}$\,\, & \,$l$\, & $n L_{\widetilde{j}}\,; \,(n-1) L_{\widetilde{j}}$ & $\widetilde{E}_{n k}\,(fm^{-1})$ \\
\hline\hline
$1$ & $-4$ & $4$ & $\frac{7}{2}$ & $3$ & $1f_{7/2}$ & \,\,$-4.489788121$\,\, & $2$ & $-4$ & $4$ & $\frac{7}{2}$ & $3$ & $2f_{7/2}$ & $\,\,-4.614788121\,\,$  \\
$1$ & $-3$ & $3$ & $\frac{5}{2}$ & $2$ & $1d_{5/2}$ & $-2.042875755$ & $2$ & $-3$ & $3$ & $\frac{5}{2}$ & $2$ & $2d_{5/2}$ & $-2.167875755$ \\
$1$ & $-2$ & $2$ & $\frac{3}{2}$ & $1$ & $1p_{3/2}$ & $-0.574728336$ & $2$ & $ -2$ & $2$ & $\frac{3}{2}$ & $1$ & $2p_{3/2}$ & $-0.699728336$\\
$1$ & $-1$ & $1$ & $\frac{1}{2}$ & $0$ & $1s_{1/2}$ & $-0.085345864$ & $2$ & $ -1$ & $1$ & $\frac{1}{2}$ & $0$ & $2s_{1/2}$ & $-0.210345864$\\
$1$ & $2$ & $1$ & $\frac{3}{2}$ & $2$ & $0d_{3/2}$ & $-4.489788121$ & $2$ & $2$ & $1$ & $\frac{3}{2}$ & $2$ & $1d_{3/2}$ & $-4.614788121$\\
$1$ & $3$ & $2$ & $\frac{5}{2}$ & $3$ & $0f_{5/2}$ & $-7.915465434$ & $2$ & $3$ & $2$ & $\frac{5}{2}$ & $3$ & $1f_{5/2}$ & $-8.040465434$\\
$1$ & $ 4$ & $3$ & $\frac{7}{2}$ & $4$ & $0g_{7/2}$ & $-12.31990769$ & $2$ & $ 4$ & $3$ & $\frac{7}{2}$ & $4$ & $1g_{7/2}$ & $-12.44490768$\\
$1$ & $ 5$ & $4$ & $\frac{9}{2}$ & $5$ & $0h_{9/2}$ & $-17.70311488$ & $2$ & $ 5$ & $4$ & $\frac{9}{2}$ & $5$ & $1h_{9/2}$ & $-17.82811488$\\
\hline 
\end{tabular}
\label{tab1} 
\end{table}

In order to have a deeper understanding of the calculations related to the energy spectrum \eqref{bs2}, we observe its values in terms of the quantum numbers $n$ and $k$ as given in Tab. \ref{tab1}. As mentioned in the previous section, in this paper we are dealing with pseudo-spin symmetry merely for gapped graphene, so that with the help of Eq. \eqref{psespinsym}, we can obtain a relationship between $k$, $\widetilde{j}$, $l$, and $\widetilde{l}$ in the states with aligned spin ($k < 0$) and unaligned spin ($k > 0$) in Tab. \ref{tab1} as well as we can write down the corresponding orbitals for $L = 0,1,2,3,4,5,...$ with shapes $s,p,d,f,g,h,...$ respectively as $n L_{\widetilde{j}}$ when $k > 0$ and $(n-1) L_{\widetilde{j}}$ when $k < 0$. Tab. \ref{tab1} shows us that there is a degeneracy energy for $k \rightarrow -k-2$, so we have $\widetilde{E}_{nk} = \widetilde{E}_{n \bar{k}-2}$ in which symbol $\bar{k}$ is $-k$. Also, we can rewrite down Eq. \eqref{bs2} in a more compact form as follows:
\begin{equation}\label{bs3}
\widetilde{E}_{n k} = A n(n+1) + B (k+1)^2 + C,
\end{equation}
where the coefficients $A$, $B$, and $C$ are
\begin{subequations}\label{coefs2}
\begin{eqnarray}
&A =  \frac{\gamma^2}{2 V_1},\label{coefs2-1}\\
&B =  \frac{\gamma^2}{2 V_1} \left(2 c_0-\frac{b_0 c_1}{b_1}-\frac{b_0 c_2}{b_2}\right),\label{coefs2-2}\\
&C =  \frac{\gamma^2}{8 V_1} \left(\frac{b_0 a_2^2}{2 b_1}+ \frac{b_0 a_2(a_1-2)}{b_1}- 2 a_0 a_2-a_1^2-a_2^2+2 a_1\right) -  \frac{\gamma^2 a_1}{4 V_2} \left(a_0+a_2\right) + C_{ps} + \widetilde{m}.\label{coefs2-3}
\end{eqnarray}
\end{subequations}

\begin{figure}[ht]
\begin{center}
{\includegraphics[scale=.3]{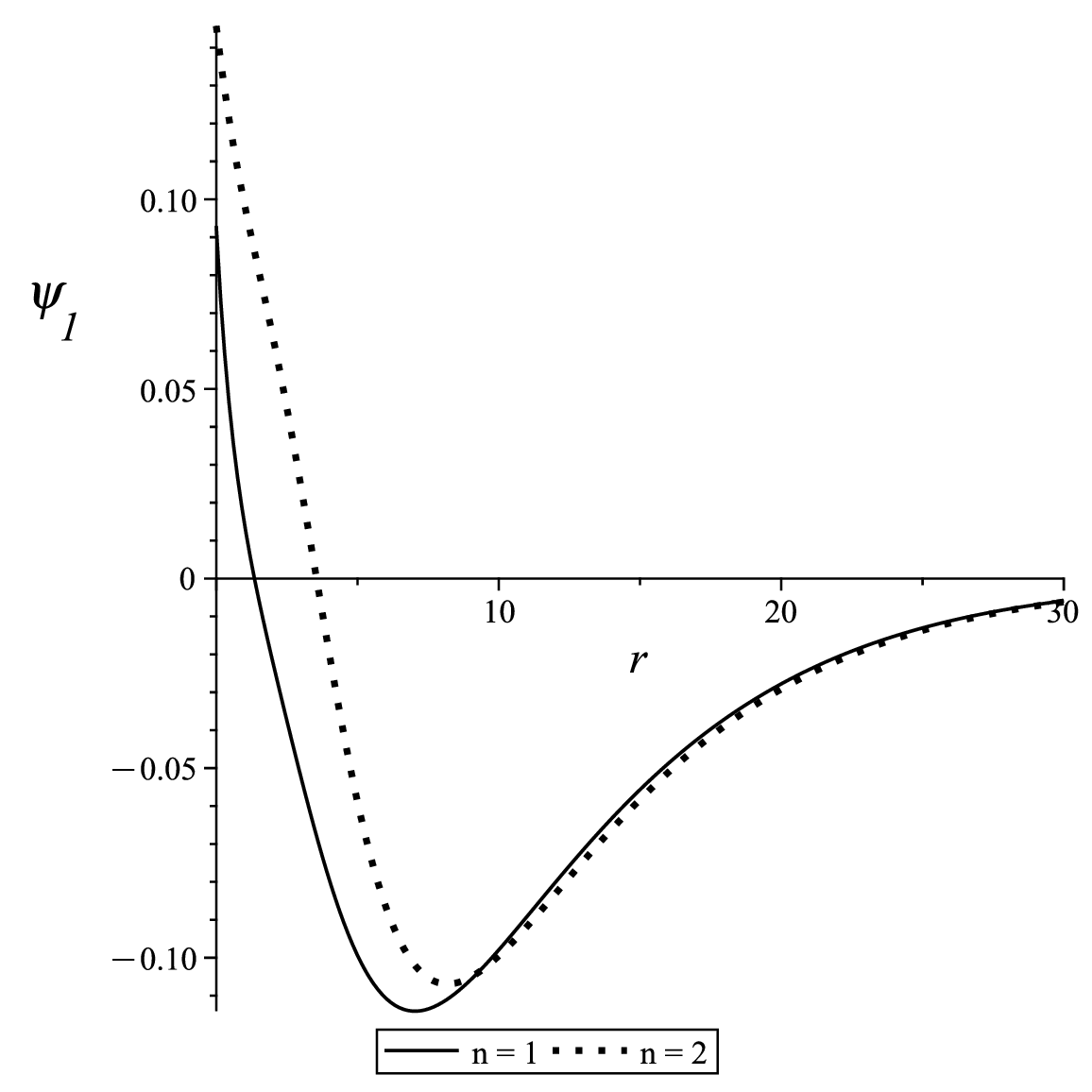}}
{\includegraphics[scale=.3]{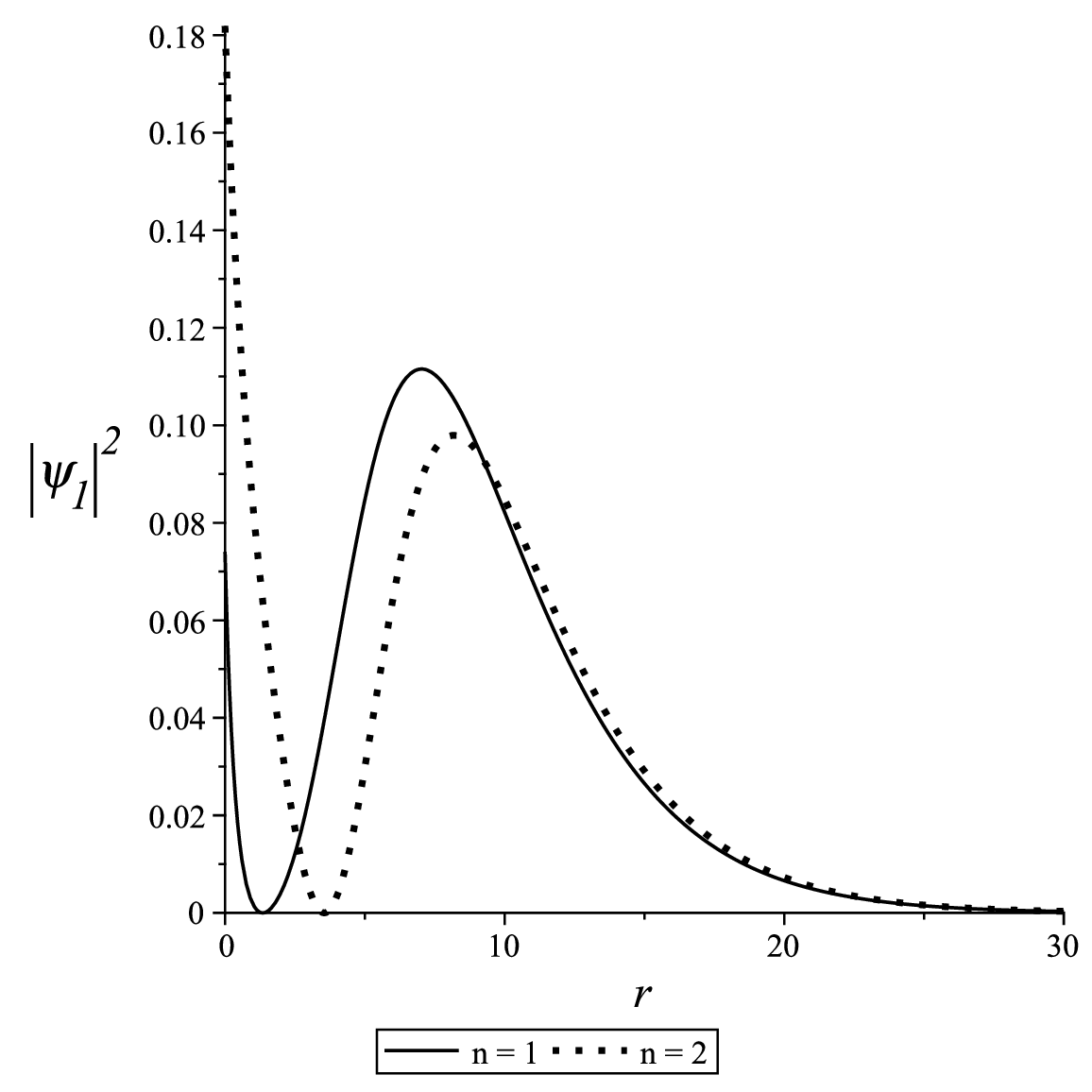}}
{\includegraphics[scale=.3]{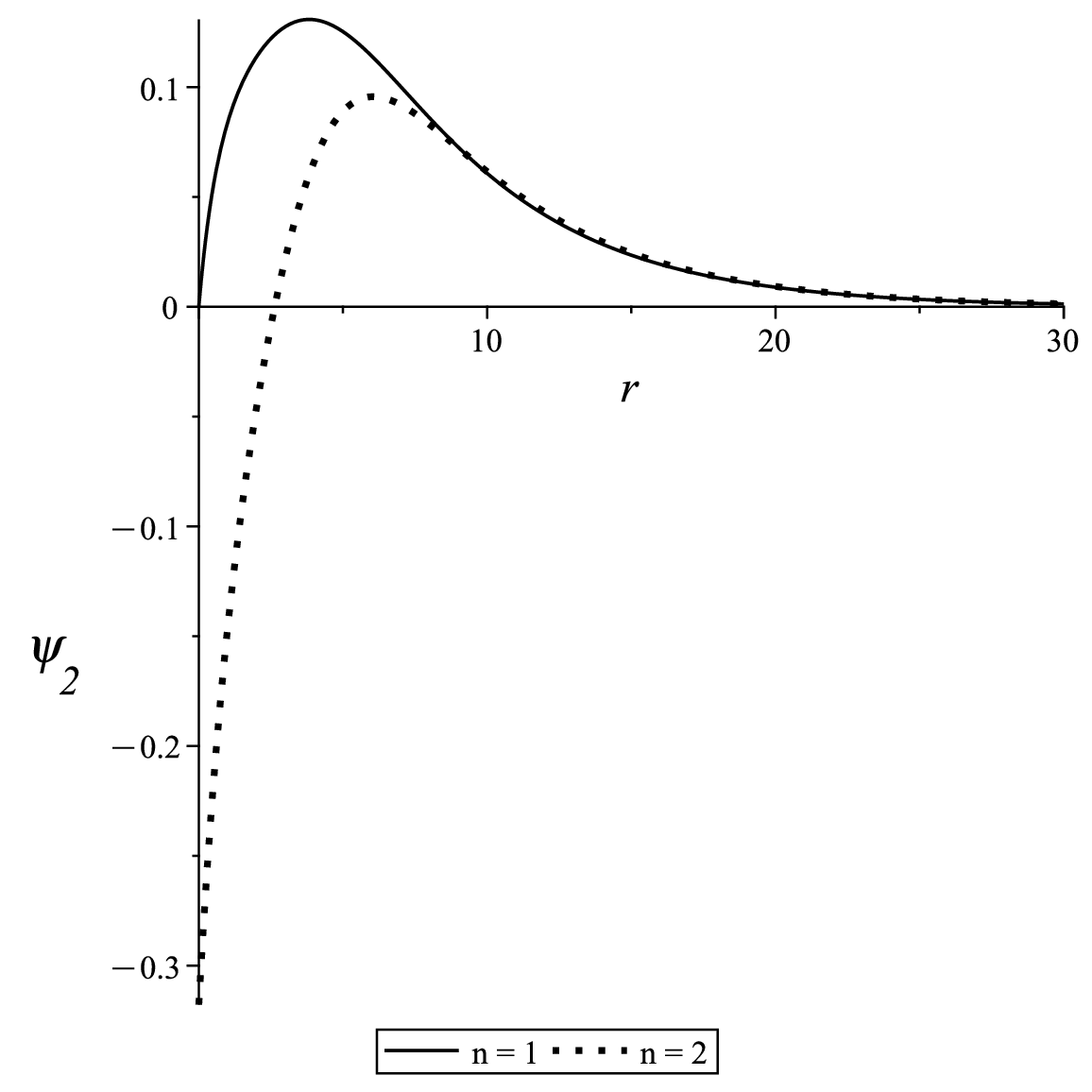}}
{\includegraphics[scale=.3]{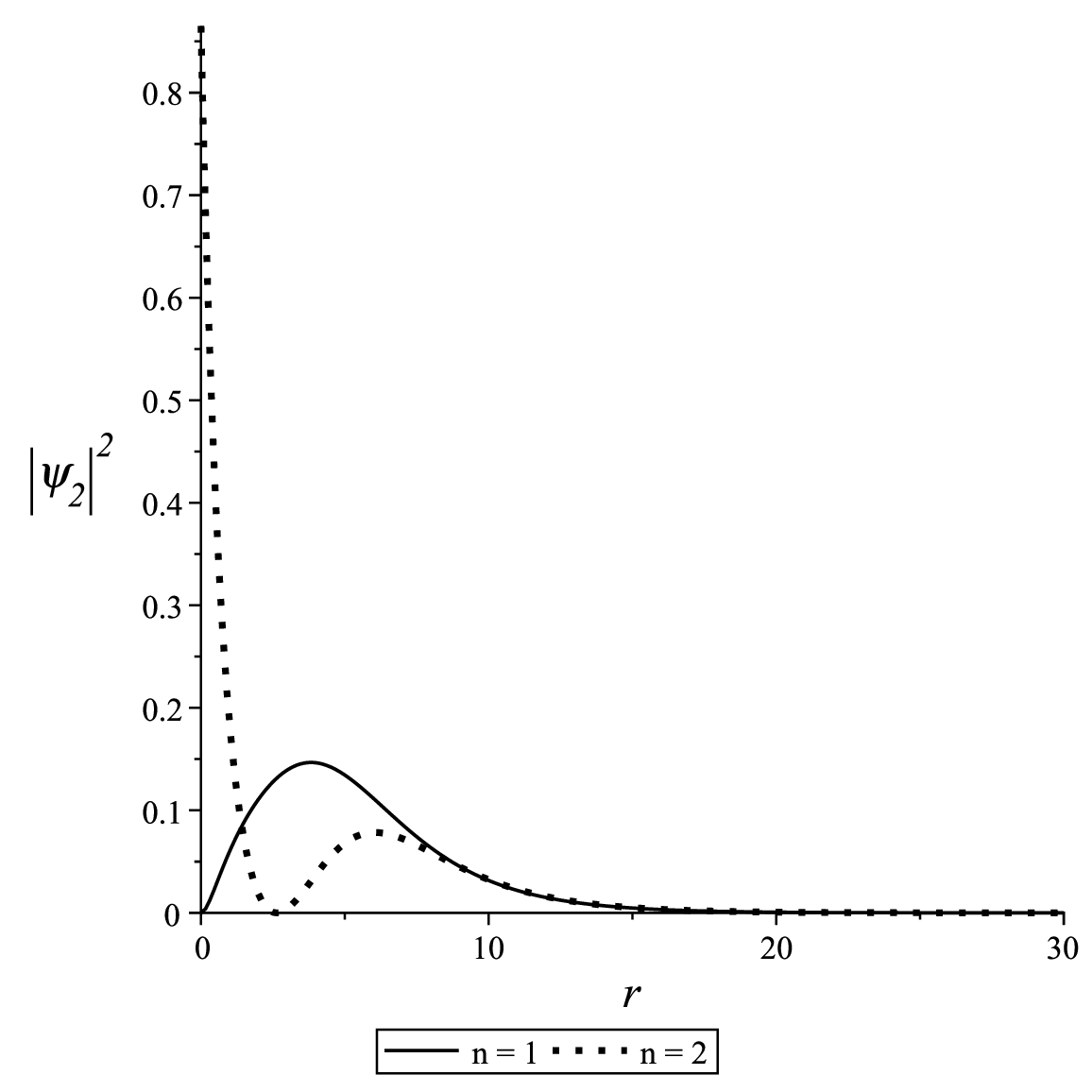}}
\caption{The real part of the wave functions $\psi_1$ and $\psi_2$ and their probabilities in terms of  coordinate $r$ in which the ground state and the first excited state are represented by the line and the dot, respectively.}\label{wavefunctions}
\end{center}
\end{figure}

In order to represent the variation of the wave functions $\psi_1$ and $\psi_2$ in terms of coordinate $r$, we first obtain wave function $\psi_2$ by substituting Eq. \eqref{R1} into Eq. \eqref{psi2} provided that $z = \tanh(\gamma r)$, and we then earn wave function $\psi_1$ by substituting the obtained $\psi_2$ into Eq. \eqref{psi12II-1}. In that case, we can see the variation of the radial part of the wave functions $\psi_1$ and $\psi_2$, and the corresponding probabilities in terms of radial coordinate $r$ for the principal quantum number $n$ in Fig. \ref{wavefunctions}. Fig. \ref{wavefunctions} depicts us that the wave functions $\psi_1$ and $\psi_2$ tend to zero when the coordinate $r$ tends to infinity.

\section{Band structure of gapped graphene}\label{IV}

From the perspective of relativistic quantum mechanics, the band structure of gapped graphene is an interesting topic in condensed matter physics and even in nanotechnology. As we know, graphene composed of carbon atoms in a hexagonal honeycomb lattice is considered as a two-dimensional material that has a unique electronic property of having a zero-band gap at the Dirac points, i.e., where the valence and conduction bands are in touch with each other. Therefore, due to the aforementioned characteristics, graphene is classified as a semi-metal with high electrical properties. On the other hand, in a more realistic investigation, especially in transistors, solar cells, and sensors, we find that graphene plays the role of a semiconductor, that is, a band gap creates between the valence and conduction bands, which is caused by factors such as defects, electric fields, and or the multi-layeredness of graphene. Therefore, the energy band gap is between the conduction and valence bands, where the conduction energy band corresponds to the energy of the electrons formed in the innermost shell of the atom, and the valence energy band corresponds to the energy of the valence electrons, which are located in the outermost part of the atomic shell.

Now, in order to explore the present work from the point of view of band structure, we note that momentum is a function of the two-dimensional wave vectors $K_x$ and $K_y$, written as $\vec{p}=\hbar K_x \vec{i}+\hbar K_y \vec{j}$. In that case, by inserting the current momentum and Eq. \eqref{magvecpot1} into Eq. \eqref{diraceq2}, we obtain the dispersion relation for gapped graphene in the presence of the Rosen-Morse potential and magnetic field as follows:
\begin{equation}\label{diraceq3}
 \left(\widetilde{E}-\widetilde{m}-C_{ps}\right)\left(\widetilde{E}+\widetilde{m}-W\right) = K_x^2+K_y^2+\eta^2 r^2+2 \eta \left(y K_x-x K_y\right),
\end{equation}
where $W$ is the same the Rosen-Morse potential, \eqref{RS1}. Therefore, in this work, the modified dispersion relation is obtained in the presence of scalar, vector and magnetic vector potentials. However, we note that if we ignore the effects of scalar and vector potentials and even the effective mass of electrons, graphene behaves as massless fermions, in which case it will have a zero-energy gap. In general, here is a more complicated problem to calculate the dispersion relation of gapped graphene in the presence of scalar and vector potentials than in the potential-free case, which must be calculated through wave vectors. In this case, we expect the dispersion relation to provide insight into the behavior of waves within material.

In what follows, we intend to represent band structure of gapped graphene in terms of wave vectors $K_x$ and $K_y$. For this purpose, we will have that the edges of the first Brillouin zone are placed with a distance in the conduction and valence bands from each other at six points, which is so-called the gap energy of graphene, and for this reason, the corresponding graphene is called gapped graphene. To show the graphene lattice vectors, we consider it as a honeycomb structure in the form of triangular lattices in a two-dimensional plane with a base of two atoms per unit cell. In this case, the network vectors, $a_1$ and $a_2$ with the lattice constant $a_0=1.42 \AA$ are as follows \cite{Neto-2009}:
\begin{subequations}
\begin{eqnarray}\label{unitcell1}
a_{1} = \frac{a_0}{2} (3, + \sqrt{3}),\label{unitcell1-1}\\
a_{2} = \frac{a_0}{2} (3, - \sqrt{3}),\label{unitcell1-2}
\end{eqnarray}
\end{subequations}
and the reciprocal network vectors $b_1$ and $b_2$ are written in the form
\begin{subequations}
\begin{eqnarray}\label{unitcell2}
b_{1} = \frac{2 \pi}{3 a_0} (1, + \sqrt{3}),\label{unitcell2-1}\\
b_{2} = \frac{2 \pi}{3 a_0} (1, - \sqrt{3}),\label{unitcell2-2}
\end{eqnarray}
\end{subequations}
also, the coordinates of the Dirac six-point wave vectors $K$ and $K'$ read
\begin{subequations}
\begin{eqnarray}\label{wavevectors1}
&K = \left(+ \frac{2 \pi}{3 a_0}, \frac{2 \pi}{3 \sqrt{3} a_0}\right), \left(- \frac{2 \pi}{3 a_0}, \frac{2 \pi}{3 \sqrt{3} a_0}\right), \left(0, +\frac{4 \pi}{3 \sqrt{3} a_0}\right),\label{wavevectors1-1}\\
&K' = \left(+ \frac{2 \pi}{3 a_0}, -\frac{2 \pi}{3 \sqrt{3} a_0}\right), \left(- \frac{2 \pi}{3 a_0}, -\frac{2 \pi}{3 \sqrt{3} a_0}\right), \left(0, -\frac{4 \pi}{3 \sqrt{3} a_0}\right).\label{wavevectors1-2}
\end{eqnarray}
\end{subequations}

\begin{figure}[ht]
\begin{center}
{\includegraphics[scale=.3]{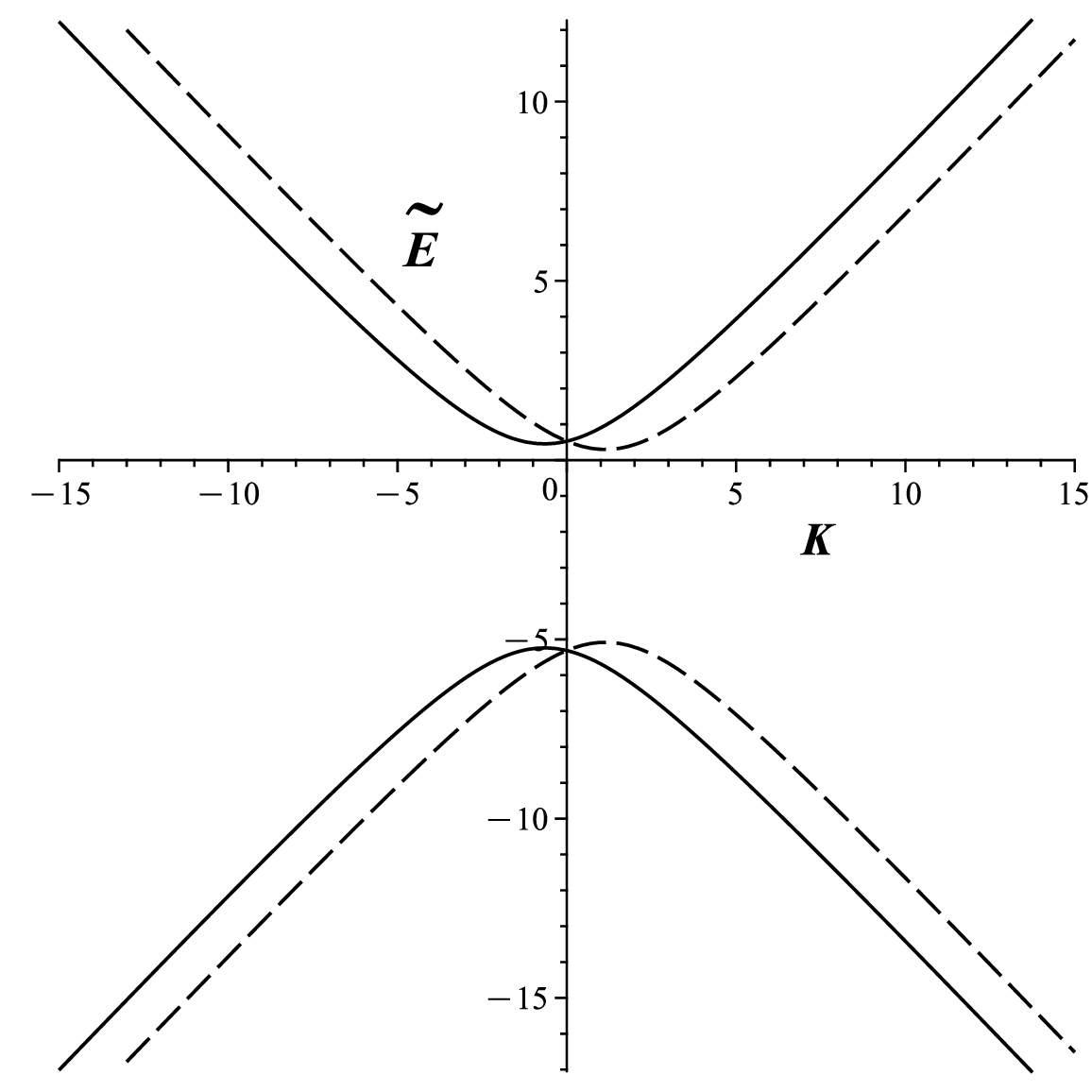}}~~~
{\includegraphics[scale=.3]{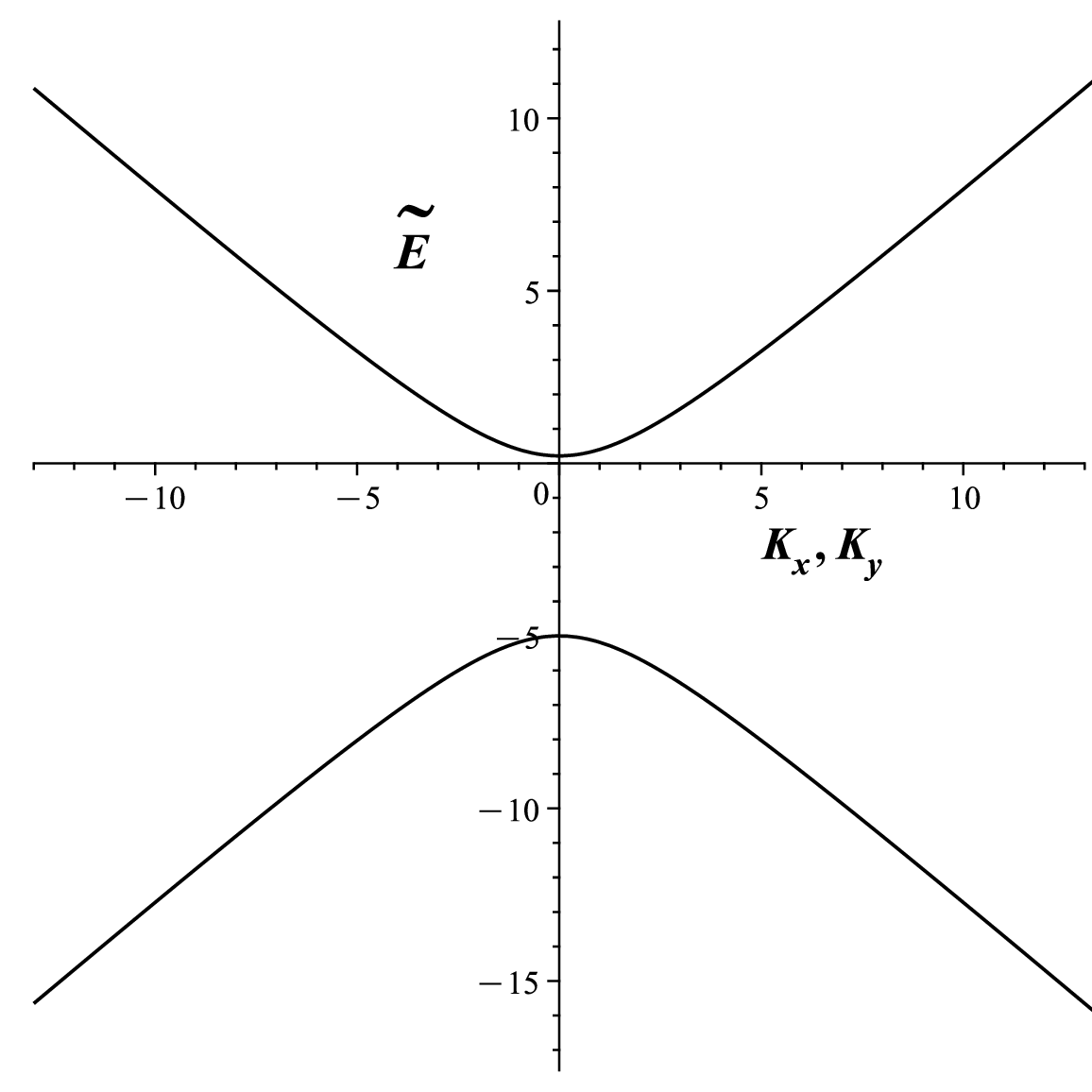}}
\caption{Left panel: the energy bands in terms of wave vectors $K_x$ (line) and $K_y$ (dash) for $\eta=1$. Right panel: the energy bands in terms of wave vectors $K_x$ and $K_y$ for $\eta=0$, which are the same.}\label{dispesion}
\end{center}
\end{figure}

However, we plot the graph of the dispersion relation \eqref{diraceq3} in terms of $K_x$ and $K_y$ for magnetic term values of $\eta=1$ (left panel) and $\eta=0$ (right panel) based on the values listed in Tab. \ref{tab1}, as shown in Figs. \ref{dispesion}. In the left panel of Figs. \ref{dispesion}, we can see that there are two energy bands as the valence band and the conduction band, and the difference between the minimum value of the conduction band and the maximum value of the valence band, which is called the gap energy, is about $5.7~ fm^{-1}$ for the component $x$, and is about $5.4~fm^{-1}$ for component $y$. We note that the size of the band gap energy depends on the mass, potential and magnetic term applied to it. Another point in Figs. \ref{dispesion} is that in both figures we see a deviation from the origin due to magnetic spins caused by magnetic fields. This means that if we omit the effect of the magnetic field into the dispersion relation \eqref{diraceq3}, so we will have a the symmetric dispersion relation in terms of wave vectors $K_x$ and $K_y$, as shown in the right panel of Fig. \ref{dispesion}. Therefore, in the modified dispersion relation \eqref{diraceq3}, when the effect of mass, scalar and vector potentials, and the external magnetic field are omitted in the Hamiltonian of the present work, the gapped graphene structure becomes a virgin graphene, in which case the band gap energy will be zero. We also observe that before and after the extremum points in the graph of the energy bands have a linear relationship with the wave vectors, which is one of the important features of graphene.

\section{Conclusion}\label{V}
In this paper, we studied the gapped graphene structure in the context of the Dirac Hamiltonian containing terms such as mass, scalar and vector potentials, and the external uniform magnetic field. The corresponding Dirac equation has explored in two-dimensional space and wrote down by the wave functions as the two-component spinors, so that they expressed in terms of polar coordinate $r-\phi$ by an arbitrary spin-orbit quantum number $k$. Since the relativity Dirac equation is related to the motion of spin half particle as fermionic quasi-particles, so it can describe the gapped graphene with the pseudo-spin symmetry approach as $k>0$ for aligned spin and $k<0$ for unaligned spin. Next, we considered the pseudo-spin symmetry as the sum of scalar and vector potentials equal to a constant value, which is written as $U(r)=C_{ps}$.

Next, to solve the current system, we converted the Dirac Hamiltonian from polar coordinates $r-\phi$ to Cartesian coordinates $x-y$ and then obtained two second-order differential equations in terms of the radial part of the wave functions $\psi_1$ and $\psi_2$. In order to describe the relativistic electron propagation in graphene under an interatomic potential, we considered the difference between the scalar and vector potentials as the Rosen-Morse potential \eqref{RS1} with bond length $r_e = -\frac{1}{\gamma} \mathrm{arctanh}\left(\frac{V_2}{2 V_1}\right)$. Although this choice makes the present study much more difficult and complicated, it is expected to have a good achievement for the construction of graphene-based electronic devices. Afterward, we have put the Rosen-Morse potential into the second-order differential equation for the wave function $\psi_2$, and then deformed it by changing the variable $z=\tanh(\gamma r)$ to reach an analytical solution. Then, we expand the trigonometric expressions with a good approximation by Taylor expansion around the point of bond length $z_e=\tanh(\gamma r_e)$ up to the second order. In the next step, with the help of the separation method, we consider that the wave function is equal to the product between the arbitrary function, $R(z)$, and the Legendre polynomial, $P_n (z)$. Then, we acquired the eigenvector \eqref{R1} and the eigenvalues \eqref{bs3} by comparing between the second and third terms of the second-order differential equations of the wave function $\psi_2$ and  the Legendre polynomial. We note that the eigenvalues are same the bound states or energy spectrum, which depends on the Rosen-Morse potential and the external magnetic potential. Next, the energy spectrum has been calculated in terms of quantum numbers $n$ and $k$ as aligned-unaligned spins with the listed values as shown in Tab. \ref{tab1} as well as we represented the corresponding calculations for orbitals with shapes $s$, $p$, $d$, $f$ , $g$, and $h$. We observe that there is a degeneracy energy for $k \rightarrow -k-2$ as $\widetilde{E}_{nk} = \widetilde{E}_{n \bar{k}-2}$ as shown in Tab. \ref{tab1}. Therefore, we plotted the graph of the wave functions $\psi_1$ and $\psi_2$ and their probabilities in terms of radial coordinate $r$ for orbitals $s$ and $p$ which are same the ground state and the first excited state, respectively.

In what follows, we studied the topology of the gapped graphene band structure using the dispersion relation. For this purpose, we first obtained the modified dispersion relation from the corresponding Dirac equation in terms of the two-dimensional wave vectors $K_x$ and $K_y$, and in this way we were able to describe the conduction and valence bands in this job. We note that the modified dispersion relation depends on the electron mass, the Rosen-Morse potential, and the external magnetic field. Finally, we plotted the variation of the conduction and valence bands in terms of wave vectors $K_x$ and $K_y$ and then calculated the corresponding energy gap or band gap for different effects of the external magnetic field. The remarkable thing in this research was that there is a curve around the minimum and maximum points of the energy bands, but beyond that, a linear relationship with the wave vector is observed.



\begin{thebibliography}{99}

\bibitem{Khveshchenko-2009}
Khveshchenko, D. V. 2009
Journal of Physics: Condensed Matter, 21(7), 075303.

\bibitem{Ulstrup-2014}
Ulstrup, S., Johannsen, J.C., Cilento, F., Miwa, J.A., Crepaldi, A., Zacchigna, M., Cacho, C., Chapman, R., Springate, E., Mammadov, S. and Fromm, F., 2014. 
Physical review letters, 112(25), p.257401.

\bibitem{Setare-2010}
Setare, M.R. and Jahani, D., 2010. 
Physica B: Condensed Matter, 405(5), pp.1433-1436.

\bibitem{Jian-2003}
Guo, J.Y., Meng, J., and Xu F.X, 2003. 
Chinese physics letters, 20(5), p.602.

\bibitem{Ikhdair-2010}
Ikhdair, S.M. and Sever, R., 2010. 
Applied Mathematics and Computation, 216(3), pp.911-923.

\bibitem{Guo-2005}
Guo, J.Y. and Sheng, Z.Q., 2005. 
Physics Letters A, 338(2), pp.90-96.

\bibitem{Sari-2015}
Sari, R.A., Suparmi, A. and Cari, C., 2015. 
Chinese Physics B, 25(1), p.010301.

\bibitem{Wei-2009}
Wei, G.F. and Dong, S.H., 2009. 
Europhysics Letters, 87(4), p.40004.

\bibitem{Wei-2008}
Wei, G.F. and Dong, S.H., 2008. 
Physics Letters A, 373(1), pp.49-53.

\bibitem{Jia-2009}
Jia, C.S., Liu, J.Y., Wang, P.Q. and Lin, X., 2009. 
International Journal of Theoretical Physics, 48, pp.2633-2643.

\bibitem{Rosen-1932}
Rosen, N. and Morse, P.M., 1932. 
Physical Review, 42(2), p.210.

\bibitem{Ikhdair-2013}
Ikhdair, S.M. and Hamzavi, M., 2013. 
Chinese Physics B, 22(4), p.040302.

\bibitem{Oyewumi-2010}
Oyewumi, K.J. and Akoshile, C.O., 2010. 
The European Physical Journal A, 45(3), pp.311-318.

\bibitem{Gang-2004}
Chen, G., Chen, Z.D., and Lou, Z.M., 2004. 
Chinese Physics, 13(3), p.279.

\bibitem{Morse-1929}
Morse, P.M., 1929. 
Physical review, 34(1), p.57.

\bibitem{Zhang-2016}
Zhang, P., Long, H.C. and Jia, C.S., 2016. 
The European Physical Journal Plus, 131(4), p.117.

\bibitem{Berkdemir-2006}
Berkdemir, C., 2006. 
Nuclear Physics A, 770(1-2), pp.32-39.

\bibitem{Ikhdair-2011}
Ikhdair, S.M., 2011. 
Journal of Mathematical Physics, 52(5).

\bibitem{Bayrak-2007}
Bayrak, O. and Boztosun, I., 2007. 
Journal of Physics A: Mathematical and Theoretical, 40(36), p.11119.

\bibitem{Amani-2012}
Amani, A.R. and Ghorbanpour, H., 2012. 
Acta Physica Polonica-Series B Elementary Particle Physics, 43(9), p.1795.

\bibitem{Hecht-1969}
Hecht, K.T. and Adler, A., 1969. 
Nuclear Physics A, 137(1), pp.129-143.

\bibitem{Arima-1969}
Arima, A., Harvey, M. and Shimizu, K., 1969.
Phys. Lett. B 30, 517 (1969).

\bibitem{Ginocchio-2004}
Ginocchio, J.N., Leviatan, A., Meng, J. and Zhou, S.G., 2004. 
Physical Review C, 69(3), p.034303.

\bibitem{Smith-1971}
Smith, G.B. and Tassie, L.J., 1971. 
Annals of Physics, 65(1), pp.352-360.

\bibitem{Bell-1975}
Bell, J.S. and Ruegg, H., 1975. 
Nuclear Physics B, 98(1), pp.151-153.

\bibitem{Jiaa-2009}
Jia, C.S., Chen, T. and Cui, L.G., 2009. 
Physics Letters A, 373(18-19), pp.1621-1626.

\bibitem{Zali-2021}
Zali, Z., Amani, A., Sadeghi, J. and Pourhassan, B., 2021. 
Physica B: Condensed Matter, 614, p.413045.

\bibitem{Wei-2010}
Wei, G.F. and Dong, S.H., 2010. 
Physica Scripta, 81(3), p.035009.

\bibitem{WeiS-2010}
Wei, G.F. and Dong, S.H., 2010. 
The European Physical Journal A, 46(2), pp.207-212.

\bibitem{Qiang-2012}
Qiang, W.C., Sun, G.H. and Dong, S.H., 2012. 
Annalen der Physik, 524(6‐7), pp.360-365.

\bibitem{Pedersen-2009}
Pedersen, T.G., Jauho, A.P. and Pedersen, K., 2009. 
Physical Review B, 79(11), p.113406.

\bibitem{Zhu-2009}
Zhu, W., Wang, Z., Shi, Q., Szeto, K.Y., Chen, J. and Hou, J.G., 2009. 
Physical Review B, 79(15), p.155430.

\bibitem{Klimchitskaya-2017}
Klimchitskaya, G.L., Mostepanenko, V.M. and Petrov, V.M., 2017. 
Physical Review B, 96(23), p.235432.

\bibitem{Novoselov-2004}
Novoselov, K.S., Geim, A.K., Morozov, S.V., Jiang, D.E., Zhang, Y., Dubonos, S.V., Grigorieva, I.V. and Firsov, A.A., 2004. 
Science, 306(5696), pp.666-669.	

\bibitem{Neto-2009}
Neto, A.C., Guinea, F., Peres, N.M., Novoselov, K.S. and Geim, A.K., 2009. 
Reviews of modern physics, 81(1), p.109.

\bibitem{Nair-2008}
Nair, R.R., Blake, P., Grigorenko, A.N., Novoselov, K.S., Booth, T.J., Stauber, T., Peres, N.M. and Geim, A.K., 2008. 
Science, 320(5881), pp.1308-1308.
  
\bibitem{Chenaghlou-2021} 
Chenaghlou, A., Aghaei, S. and Ghadirian Niari, N., 2021. 
The European Physical Journal D, 75(4), p.139.

\bibitem{Onyenegecha-2021} 
Onyenegecha, C.P., Opara, A.I., Njoku, I.J., Udensi, S.C., Ukewuihe, U.M., Okereke, C.J. and Omame, A., 2021. 
Results in Physics, 25, p.104144.

\bibitem{Eshghi-2016} 
Eshghi, M. and Mehraban, H., 2016. 
Journal of Mathematical Physics, 57(8).

\bibitem{Alimohammadian-2020} 
Alimohammadian, M. and Sohrabi, B., 2020. 
Scientific reports, 10(1), pp.1-10.

\bibitem{Downing-2016}
Downing, C.A. and Portnoi, M.E., 2016. 
Physical Review B, 94(16), p.165407.

\bibitem{Gupta-2008}
Gupta, K.S. and Sen, S., 2008. 
Physical Review B, 78(20), p.205429.

\bibitem{Hassanabadi-2012}
Hassanabadi, H., Maghsoodi, E., Oudi, R., Zarrinkamar, S. and Rahimov, H., 2012. 
The European Physical Journal Plus, 127(10), p.120.

\bibitem{Arda-2015}
Arda, A. and Sever, R., 2015. 
Communications in Theoretical Physics, 64(3), p.269.

\bibitem{Min-2008}
Min, H., Borghi, G., Polini, M. and MacDonald, A.H., 2008. 
Physical Review B, 77(4), p.041407.

\bibitem{Jose-2009}
San-Jose, P., Prada, E., McCann, E. and Schomerus, H., 2009. 
Physical review letters, 102(24), p.247204.

\bibitem{Tuan-2014}
Tuan, D.V., Ortmann, F., Soriano, D., Valenzuela, S.O. and Roche, S., 2014. 
Nature Physics, 10(11), pp.857-863.


\end{thebibliography}
\end{document}